\pdfoutput=1
\PassOptionsToPackage{dvipsnames}{xcolor}
\documentclass[11pt]{article}

\usepackage[final]{ACL2023}

\usepackage{times}
\usepackage{latexsym}
\usepackage{graphicx}
\usepackage[T1]{fontenc}
\usepackage[utf8]{inputenc}

\usepackage{microtype}

\usepackage{inconsolata}

\usepackage{hyperref}
\usepackage{makecell}
\usepackage{xurl}

\usepackage{xparse}
\definecolor{bluepigment}{rgb}{0.2, 0.2, 0.6}

\ifx\nocomment\undefined
    \NewDocumentCommand{\heng}
    { mO{} }{\textcolor{red}{\textsuperscript{\textit{Heng}}\textsf{\textbf{\small[#1]}}}}
\else
    \NewDocumentCommand{\heng}
\fi

\newcommand{\colorado}[1]{\textcolor{red}{\textbf{Susan or Martha} #1}}
\newcommand{\zoey}[1]{\textcolor{blue}{\textbf{Zoey}: #1}}

\newcommand{\add}[1]{\textcolor{red}{\textbf{[add cite: #1]} }}

\newcommand{\interface}{\url{https://www.kairos.jiaxuan.me}}
\newcommand{\video}{\url{https://www.youtube.com/watch?v=myru-fozVWI}}

\title{Human-in-the-Loop Schema Induction}

\author{
  Tianyi Zhang$^*$$^1$, 
  Isaac  Tham$^*$$^1$, 
  Zhaoyi Hou$^*$$^1$, 
  Jiaxuan Ren$^1$, 
  Liyang Zhou$^1$\\
  {\bf Hainiu Xu$^1$, 
   Li Zhang$^1$, Lara J. Martin$^1$, 
  Rotem Dror$^1$,
  Sha Li$^2$, Heng Ji$^2$}\\
  {\bf Martha Palmer$^3$, Susan Brown$^3$, Reece Suchocki$^3$, and
  Chris Callison-Burch$^1$}\\ 
  $^1$ University of Pennsylvania, $^2$ University of Illinois Urbana-Champaign\\$^3$ University of Colorado, Boulder,  $^*$ equal contribution\\
  \{zty, joeyhou, ccb\}@upenn.edu
  }

\begin{document}
\maketitle

\begin{abstract}
Schema induction builds a graph representation explaining how events unfold in a scenario. 
%Many approaches to schema induction have been developed under the DARPA KAIROS program and similar ones.
% \rotem{I don't think that the name of the project is relevant, because other methods could have been developed under different projects/researches}. 
Existing approaches have been based on information retrieval (IR) and information extraction (IE), often with limited human curation.  We demonstrate a human-in-the-loop schema induction system powered by GPT-3.\footnote{Webpage: \interface;\\Video:\video} We first describe the different modules of our system, including prompting to generate schematic elements, manual edit of those elements,
% \rotem{this sounds a bit like you are selecting humans and not that humans perform the selection, maybe write "manual selection process"}, 
and conversion of those into a schema graph.  By qualitatively comparing our system to previous ones, we show that our system not only transfers to new domains more easily than previous approaches, but also reduces efforts of human curation thanks to our interactive interface.

\end{abstract}

\section{Introduction}
% definition of schema induction and current approaches of schema induction (gpt3)
% what event schema induction is and why this is important
% 1. event schema and significance of event schema
% 2. event schema induction
% 3. current approach
% 4. our approach

% 1. event centric nlp and event schema - why important? because they have lots of applications!
%TODO 1: citation at the end of the paragraph
Event-centric natural language understanding (NLU) has been increasingly popular in recent years. Systems built from an event-centric perspective have resulted in impressive improvements in numerous tasks, including open-domain question answering \cite{yang_open_domain_qa}, intent prediction \cite{rashkin-etal-2018-event2mind}, timeline construction \cite{do-etal-2012-joint}, text summarization, \cite{text_summarization_citation} and misinformation detection \cite{fung-etal-2021-infosurgeon}. At the heart of event-centric NLU lie event schemas, an abstract representation of how complex events typically unfold. The study for such a representation dates back to the 70s, where scripts were proposed as a series of sequential actions \cite{script_schank}. Back then, the schemas were limited to linear and temporal ones. 
% \rotem{maybe mention it is a graph representation where the vertices are sub-events and the edges are relations between events}. 
A more recent formulation of event schemas is a graph where the vertices are event flows and the edges are temporal or hierarchical relations between those events \cite{du-etal-2022-resin}. 
% \harry{Because the above schema formulation is not the only possibility, you want to not only cite existing work, maybe Heng's papers, but also ideally touch on why this representation is good.}
% aside from event-centric methods, there were some other methods, representation of entities, 
% 
\begin{figure}
\vspace{-4mm}
\includegraphics[width=.4\textwidth]{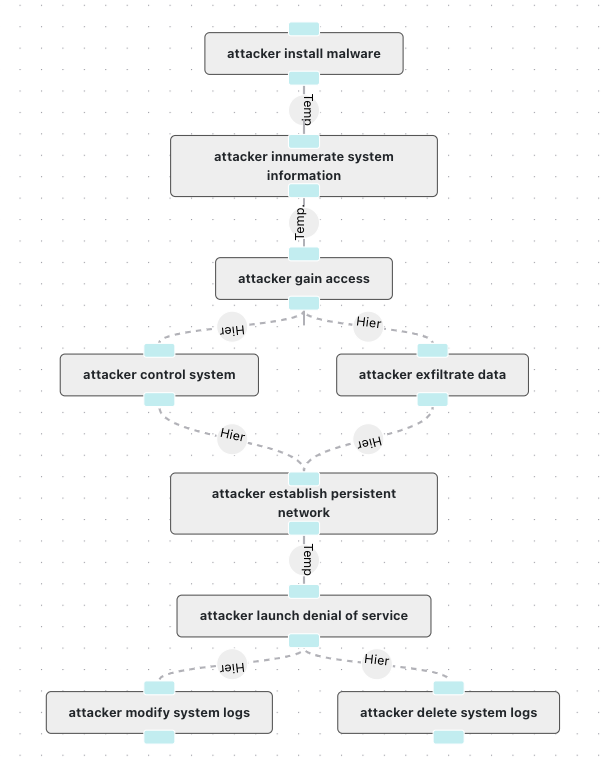}
\centering
\caption{An example of Cyber Attack schema. The tree structure represents the temporal and hierarchical relations between the nodes.}\label{schema_example}
\vspace{-4mm}
\end{figure}

For example, as shown in Figure \ref{schema_example}, the event schema for a "cyber attack" could include sub-events such as "gain access", "control system", "exfiltrate files", "modify system logs", etc. The schema would also include the relationships between these sub-events. For instance, the event "gain access" would take place \textit{before} the event "modify system logs" since a person needs access to a system before modifying it. For the same reason, "exfiltrate data" would only take place \textit{after} "gain access". Event schemas like this encode high-level knowledge about the world and allow artificial intelligence systems to reason about unseen events \cite{du-etal-2022-resin}.

\begin{figure*}[h]
\vspace{-2.5mm}
\includegraphics[width=0.75\paperwidth]{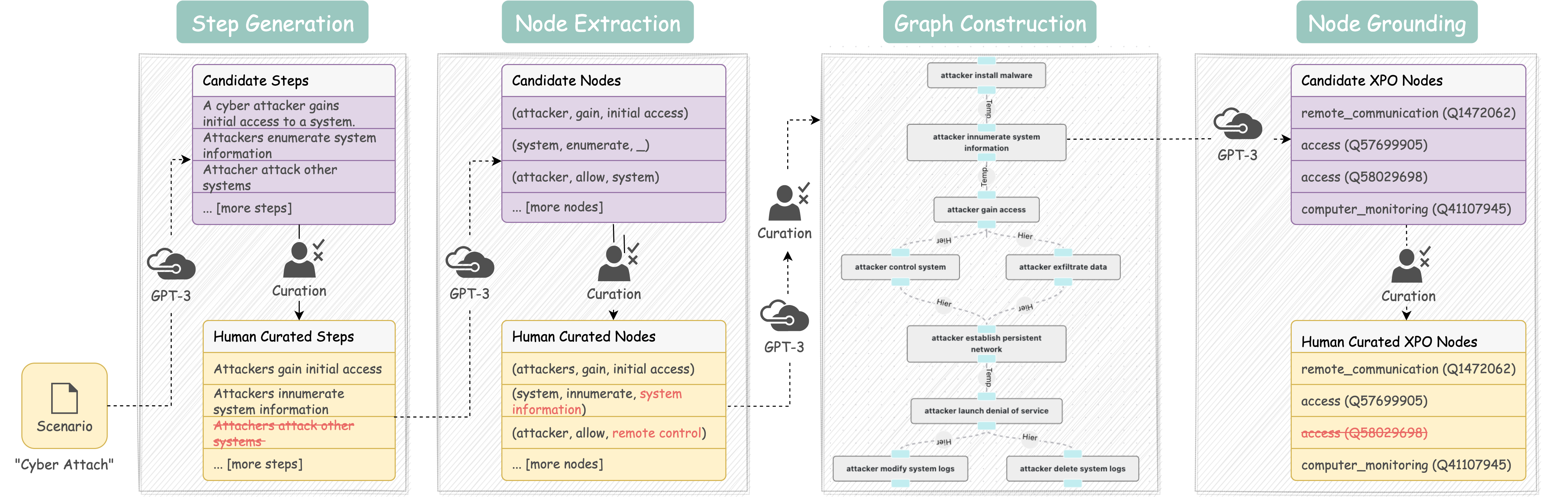}
\centering
\caption{Our schema curation system includes four main stages: Step Generation, Node Extraction, Graph Construction and Node Grounding; Model ouput is highlighted in \colorbox{Thistle}{purple background}; Human curated output is highlighted in \colorbox{Goldenrod}{yellow background}; human curation is shown in \textcolor{red}{\textbf{red}}.}
\centering
\label{system_fig}
\vspace{-2.5mm}
\end{figure*}

% 2. previous work in automatic schema induction and human-writing schemas - why hard? because it is hard!
The DARPA Knowledge-directed Artificial Intelligence Reasoning Over Schemas (KAIROS) program\footnote{\url{https://www.darpa.mil/program/knowledge-directed-artificial-intelligence-reasoning-over-schemas}} aims at developing schema-based AI systems that can identify, comprehend, and forecast complex events in a diverse set of domains. To enable such a system, scalable generation of high-quality event schemas is very crucial. On one hand, fully-manual schema creation at a large scale can be inefficient, since people have diverse views about a certain concept, leading to inconsistent schema results. On the other hand, fully automated systems are scalable, but not with high-quality.
% \tianyi{not in high quality; domain transfer problem comes from IR-IE system, not automatic schema induction, so our gpt system won't have this problem}. 
In fact, the majority of existing approaches under the KAIROS program are fully-automated IR and IE systems over large collections of news articles \cite{li2020connecting,li2021future}. 
% As fully-automated systems, all of them do not involve any human feedbacks or curations as part of the schema generation. 
Only some of limited human post-processing on schemas \cite{ciosici-etal-2021-machine} have been explored. Further discussion of the advantages and limitations of existing systems can be found in Related Work.
% %Arguably, these approaches may be limited in coverage to news-oriented events. 
% The KAIROS program also allows a small amount of human curation of the output of the automatic system. \harry{Why is this sentence here? Maybe discuss whether the previous IR IE methods use human curation at all? I assume they are fully-automated, so they don't use any curation?}

%TODO \joey{why not fully-automated? challenging, why?}
Instead of focusing on fully-automated schema induction systems, we propose a human-in-the-loop schema induction pipeline system. Rather than using IR and IE over a large document collection, our system relies on pre-trained large language models (LLMs) and human intervention to jointly produce schemas. 
% First, our interface provides a guided, interactive procedure that allows users to create schemas via GPT-3 \cite{brown2020language}.
% % \zoey{cite GPT3 paper}. 
% The guided procedure allows the user to create a structured representation from an LLM,
% % \rotem{make sure to write large language models explicitly and the LLM before the first time you use LLM}
% then select and edit its suggestions. \harry{This is not a motivation - this is what our system does.} This provides researchers the chance to review schemas and spot errors in the models suggestion. 
Our main motivation is that human-verified schemas are of higher quality. That is because human curation can filter out failure cases such as incompletness, instability, or poor domain transfer results in previous systems \cite{dror2022zero, peng2019knowsemlm}. With human curation, schemas are more reliable and accountable when applied to downstream tasks such as event prediction. This is significant if the downstream tasks involve safety-critical applications like epidemic prevention, where the quality of the schema matters beyond task performance numbers.

Figure \ref{system_fig} is a flowchart of our four-stage schema induction system: \textbf{step generation}, \textbf{node extraction}, \textbf{graph construction}, and \textbf{node grounding}. 
\iffalse
Step generation aims to create scenario steps (initially represented as natural language sentences), from a prompt. Then, the node extraction step creates simplified nodes using a subject-verb-object format that is derived from the sentences. After that, graph construction creates temporal and hierarchical links between the nodes. The last stage, node grounding, can map the nodes to an ontology, e.g., Wikidata or DARPA's Cross-Program Ontology (XPO). \colorado{Could you please add a citation for the XPO.} \zoey{Mention that this process can be done iteratively?} 

Step generation aims to create scenario steps (initially represented as natural language sentences), from a prompt. Prompts can vary – the default is to list the steps involved in a scenario. Then, the node extraction step creates simplified nodes using a subject-verb-object format that is derived from the sentences. After that, graph construction creates temporal and hierarchical links between the nodes.

At this point we have a simple but complete schema for the scenario. 
The last stage, node grounding, can map the nodes to an ontology, e.g., Wikidata or DARPA's Cross-Program Ontology (XPO). \colorado{Please add a citation for the XPO.} 
\fi
Each stage has two main components: the LLM (e.g. GPT-3) at the back-end to output predictions (the purple boxes in the figure) and an interactive interface at the front-end for human curation of the model output (the yellow boxes). The GPT-3 prompts that are used in each stage of the process are given in the Appendix \ref{sec:appendix}, along with example inputs and outputs.

A more comprehensive description of the implementation and functionalities of our interface can be found in Section \ref{sec: Implementation}. A case study is  given in Section \ref{sec: Evaluation}. It walks through each step in our pipeline system under an example scenario, cyber attack. Also, in Section \ref{sec: Evaluation}, we provide a qualitative evaluation of five example scenarios. The summary and discussion of our system are included in Section \ref{sec: Conclusion}.

% \harry{I suggest removing the contribution statements, because they're really talking about 1 thing.}
\iffalse
The contributions of our work are:
\begin{enumerate}
  \item We offer a guided, interactive approach to schema induction using large language models that allows easy human curation.
  \item We create an online web interface \footnote{\interface} to allow users to experiment with our approach.
\end{enumerate}
\fi

%TODO \joey{slightly more details about how our system work, e.g. 2-3 sentences for each component}
% evaluation and looking ahead
% \joey{evaluation metric, people involved in evaluation, feedbacks}\\
% \joey{future improvements, codebase, web-link}

\section{Related Work}
% \zoey{Try to compress this section to half a page. Some paragraphs like those on data generation are not relevant.} 
% schema induction (Tianyi&Hainiu)
% human in the loop (Joey)
% schema evaluation (Isaac&Leon)
% other schema interface (why we need to have our interface) (Joey)
\vspace{-2mm}
\subsection{Schema Induction}
\vspace{-2mm}
% Schema Induction has been studied for a long time \rotem{rephrase, or just start with the next sentence "Early work on schema induction ..."}. 
Early work from \citet{chambers2008unsupervised,chambers2009unsupervised} automatically learned a schema from newswire text based on coreference and statistical probability models. 
% For instance, a schema they learned in an `Arrest' scenario was: police search suspect → police arrest suspect → jury sentence suspect. 
Later, \citet{peng2016two,peng2019knowsemlm} generated an event schema based on their proposed semantic language model (like an RNN structure). Their work represented the whole schema as a linear sequence of abstract verb senses like {\tt arrest.01}
% {\tt arrest.01-charge.05-convict.01} 
from VerbNet \cite{schuler2005verbnet}. Those works had two main shortcomings: first, the schema was created for a single actor (protagonist), e.g. suspect. It caused limited coverage in a more complex scenario, e.g. business change-acquisition; second, the generated schema, a simple linear sequence, failed to consider different alternatives such as XOR.
% \rotem{do you mean they could not support multiple events occurring at once?}; 
% third, the coarse schema focused on triggers but ignored the interaction between different participants and their roles, e.g. the target of an attack event could be the victim of an injured event.

More recently, \citet{li2020connecting,li2021future} used transformers to handle schema generation in a complex scenario. It viewed a schema as a graph instead of a linear sequence. 
% For example, the suspect in the above example played different roles in different events, the attacker of an attack event and the detainee of an arrest event. 
However, this approach was unable to transfer to new domains where the supervised event retrieval and extraction model failed. \citet{dror2022zero} took GPT-3 generated documents to build a schema. Although it bypassed the event retrieval and extraction process and solved the domain transfer problem, it suffered from the incompleteness and instability of GPT-3 outputs.
% \zoey{There are some inaccurate characterizations of Manling's paper and Rotem's paper. For Manling's paper, there was no supervised event retrieval but a domain specific dataset. For Rotem's method, the event extraction pipeline was still a central part of our system. I also see that you emphasized entity relations, so a natural question would be: does your current model handle this? } 

Currently, neither do they offer a perfect solution for schema induction without manual post-processing, nor build a timely human correction system \cite{du-etal-2022-resin}.  
% \zoey{None of the previous papers actually mention human curation postprocessing, cite Du et al 2022 to motivate this. } 
Our demonstration system 
% is an extension of the research described in \cite{anonymous2023opendomain}.
% \add{Cite Zoey's paper again} \zoey{Change  ``is implementation of'' to ``is extension of ''} We 
develops a curation interface that can generate a comprehensive schema easily with a human curator in the loop.  The curated data collected through our tool could be useful for fine-tuning and improving the models.

\vspace{-2mm}
\subsection{Human-in-the-loop Schema Curation Interface}
\vspace{-2mm}

Another area related to our work is  human-in-the-loop schema generation, where annotators collaborate with computational models to create high-quality event schema. 
% In this field, recent works either focus on studying the computational models that help such interaction(i.e. computational models for specific data distribution) or on the interactive interfaces for such collaboration.
% For dataset generation research, recent work has shown the power of large language models such as GPT-3 for various applications, including generating challenging examples for NLI task \cite{liu2022wanli}, structural data synthesis \cite{yuan2021synthbio} and hate speech detection \cite{tekiroglu-etal-2020-generating}. These works have shown that, given proper prompts, pre-trained LLMs are able to generate targeted text for pre-defined purposes. However, the evaluation for these works mainly focused on the quantity of the dataset \rotem{unclear, what is the quantity of the dataset?} and the quality of downstream tasks, instead of the annotation costs in terms of human labor, which is one of our main focuses.
In this field, one of the closest approachs is the Machine-Assisted Script Curation \cite{ciosici-etal-2021-machine} created for script induction. 
% \zoey{also cite the schema interface from Colorado \url{https://aclanthology.org/2021.acl-demo.19.pdf}} 
% Colorado \cite{mishra2021graphical}
With a fully interactive interface, they have shown the feasibility of real\-time interaction between humans and pre-trained LLMs (e.g. GPT-2 or GPT-3). The main differences are the level of automation and adaptability to other generative models. In terms of automation, our interface makes use of pre-trained LLMs to automatically generate schema content, compared to their interface which largely counts on human input. For adaptability, our interface supports the curation of the schema generated by different language models (e.g. GPT-3 models with different sizes), which makes it possible for users to evaluate the generations of different models. In contrast, there is no such possibility in their interface. % \zoey{Is this adaptability mentioned again in other parts of the paper?}

Another interface built for schema curation focuses on visualization of the schema structure, such as the temporal relations between event nodes and internal relations among entities \cite{mishra2021graphical}. While this interface provides a user-friendly experience when it comes to schema graph curation, it requires the user to come up with the content of event schemas in json format, which requires much more human effort compared to our interface. In addition, our interface also provides an optional grounding function after the event graph curation step, which is not presented in this interface.

\section{Terminology and Problem Definition}
Our work focuses on efficiently building a schema graph of a scenario using both LLMs and human input. Following the workflow of our system (see the workflow in Figure \ref{system_fig}), a \textbf{scenario} is a general event type that an interested party
% a human \rotem{maybe instead of a human write "an interested party"} 
will build the schema for, e.g. a `disease outbreak'. \textbf{Steps} are a list of sub-events generated by GPT-3 according to a prompt in the step generation stage. Each step can be a phrase or a short sentence, such as `spread to other areas', etc. \textbf{Nodes} or \textbf{tuples} are subject-verb-object pairs extracted from steps at the node extraction stage, such as `(disease, spread, to other area)'. \textbf{Graphs} are a visualization of the schema, whose edges joining the nodes represente temporal and hierarchical relations.
% \rotem{this paragraph should refer to an example in the form of an image}

\section{Implementation}
\label{sec: Implementation}
% gpt3 prompt and completion (Tianyi)
% node extraction (Isaac)
% naming (Tianyi)
% graph building (Leon)
% grounding (Hainiu)

Our pipeline system contains four sequential stages: \textbf{step generation}, \textbf{node extraction}, \textbf{graph construction}, and \textbf{node grounding}. A flowchart of the interface system is shown in Figure \ref{system_fig}. The \textbf{step generation} stage generates steps for a scenario and the user can specify how many steps they would like to generate. The \textbf{node extraction} stage extracts nodes (subject-verb-object tuples) from the previous verbose steps. The \textbf{graph construction} stage orders the extracted nodes temporally and hierarchically. Meanwhile, modifications of the nodes are still possible. The \textbf{node grounding} stage maps node text to a node in the XPO ontology \cite{elizabeth2023darpa} (derived from WikiData\footnote{\url{https://www.wikidata.org/wiki/Wikidata:Main_Page}}). 
% \rotem{cite, add link, something}. 
The flexible interface system allows users to either go through the entire process
% \rotem{to either go through the entire process} 
to create a schema from scratch or directly start at any stage to edit the model's prediction. In addition, the back-end GPT-3 models can be replaced by other user pre-trained models if deployed locally.
% \rotem{this sentence is unclear}.

\vspace{-1mm}
\subsection{Step Generation}
\vspace{-1mm}
The step generation stage aims at generating steps given a scenario. At the backend, zero-shot GPT-3 incorporates a user's input into a prompt and generates ordered steps. The interface allows users to generate steps quickly with prompt templates\footnote{an \{event type\} appended to a predefined prompt: Before, After or Sub-steps} or finetune the generated steps with user-designed prompts. A typical use case of the user-designed prompts is to expand a certain step to more detailed steps. For instance, a template prompt "List the steps involved in \{disease outbreak\}:" may create steps such as "1. Identify the symptoms of the disease; 2. Collect data from affected individuals; ...". Then, the user can re-prompt for, e.g., the second step, “List the steps involved \{step2\} in detail:”. Additionally, users can modify and select GPT-3 generated steps easily by clicking on them. 
% The step generation interface is capable of handling a user's interaction multi-times with GPT-3 and saving all the results. 
When the 'save' button is clicked, all user selected steps will be saved in the database for the use of the node extraction stage or further fine-tuning of the step generation model. A screenshot of the step generation interface with user’s operations can be seen in Figure \ref{case_study_step}.

\iffalse

\begin{figure}[h]
\includegraphics[width=.45\textwidth]{steps.png}
\centering
\caption{Prompt-Completion Interface}
\centering
\label{steps}
\end{figure}

\subsection{Naming}
GPT-3 generated steps in the prompt-completion stage can be verbose, e.g. "A risk assessment is carried out to determine the potential public health impact of the outbreak". The naming part aims at generating a succinct phrase (name) summarizing a step, e.g. "assessment". Our GPT-3 naming model forms a few-shot prompt with the format: "Description:{step} Name:{name}" (see few-shot examples in appendix),  to generate the name for an input step. The generated name can be modified by the user then saved when clicking a 'save' button. This naming model can be fine-tuned with human curated data or replaced with other summarization models. See a screenshot of naming interface in Figure \ref{naming}.
\begin{figure}[h]
\includegraphics[width=.45\textwidth]{naming.png}
\centering
\caption{Naming Interface}
\centering
\label{naming}
\end{figure}
\fi
\vspace{-1mm}
\subsection{Node Extraction}
\vspace{-1mm}
Nodes are structured representations of events in the form of a \{subject, verb, object\} tuple. Node extraction is to extract these nodes from the GPT-3 generated steps saved in step generation stage, which are unstructured sentences.

There are two methods, based on AllenNLP \cite{shi2019simple} or GPT-3, that users can choose from to extract nodes. The former uses AllenNLP's Semantic Role Labelling (SRL) model to extract nodes from the steps. The SRL model implements a BERT \cite{devlin2018bert} sequence prediction model to identify the predicates and the arguments (e.g. A0, A1) in a text. We simply choose the identified A0 as subject, A1 as object, and predicate as the verb to form a node. An optional coreference resolution model can be used to resolve referenced entities between the different steps with an AllenNLP's SpanBERT-based model \cite{Lee2018HigherorderCR}. Here, we concatenate all the steps and replace a pronoun with its referenced entity (noun) in the original steps.

The GPT-3 node extraction method uses instructional few-shot prompting to extract \{subject, verb, object\} tuples from the steps. Several example sentences are given to show GPT-3 the expected syntactic and semantic output. We follow \cite{liu-etal-2022-makes}'s recommendation for few-shot design by including context examples that are semantically similar to the KAIROS application environment (daily life and news).
% \harry{Are the readers supposed to know what the test domain is at this point? If so, where is this info mentioned?} 
See appendix \ref{sec:appendix} for our few-shot prompts.

The extracted nodes are shown to the user in a table with 3 columns (subject, verb, object). For example, for “The CDC collects and analyzes data on disease outbreaks”, one of the extracted nodes is “The CDC (subject) collects (verb) data (object)”.
% \harry{Maybe give an example with a subject?}. 
Users are able to choose and edit nodes (tuples). User edits are saved and will be used for graph construction and fine-tuning of the GPT-3 node extraction model.

\iffalse

\begin{figure}[h]
\includegraphics[width=.45\textwidth]{event tuple.png}
\centering
\caption{Event Tuple Extraction Interface}
\centering
\label{eventtuple}
\end{figure}

\begin{figure}[h]
\includegraphics[width=.45\textwidth]{event tuple prompt.png}
\centering
\caption{GPT-3 assisted Event Tuple Extraction Prompt}
\centering
\label{eventtuple_prompt}
\end{figure}

\begin{figure}[h]
\includegraphics[width=.3\textwidth]{annotator.png}
\centering
\label{schema_example_interface}
\caption{Graph Interface}
\end{figure}

\begin{figure}[h]
\includegraphics[width=.45\textwidth]{grounding.png}
\centering
\caption{Grounding Interface}
\label{grounding}
\end{figure}\\
\fi
\vspace{-1mm}
\subsection{Graph Construction}
\vspace{-1mm}
In the graph construction stage, our system automatically adds temporal and hierarchical edges to the previously extracted nodes. The edges are created using zero-shot GPT-3 with multiple choice questions. For each pair of nodes, GPT-3 is instructed to choose between ‘Before’, ‘After’, ‘Same time’ or ‘no relation’ for temporal eges; and ‘Parent’, ‘Child’ or ‘no relation’ for hierarchical edges. For example, for the node pair “collect data” and “identify the signs and symptoms”, GPT-3 predicts ‘After’ for temporal order and ‘no relation’ for hierarchical order, in which case we will add a temporal edge from “identify the signs and symptoms” to “collect data”, and no hierarchical edge will be created. If a conflict occurs between (node1, node 2) pair and (node2, node1) pair, e.g. ‘After’ and ‘After’ for a temporal order or ‘Parent’ and ‘Parent’ for a hierarchical order, we will treat it as no relation to resolve the conflict, thus adding no new edges to the graph.

The graph construction interface allows users to modify the GPT-3 generated schema with ease. After predicting both temporal and hierarchical relations between all pairs of nodes, the interface will display the graph via the Vis-network framework\footnote{\url{https://www.npmjs.com/package/react-vis-network-graph}}. 
% \harry{Add either citation or a footnote link.}. 
It supports adding, editing, deleting graph nodes and edges. When the user clicks on a node, the detailed information including the ID and description of a node will be shown as well as the button to delete or edit the node. By clicking the edge, users can modify the edge type or delete it. Users will be able to create a new node by double clicking and a new edge by dragging and dropping an arrow from two nodes. A screenshot of our graph construction interface can be seen in figure \ref{case_study_graph}.

\vspace{-1mm}
\subsection{Node Grounding}
\vspace{-1mm}
Although a schema (graph) is completely created after the previous stages, some nodes may express the same semantic information, e.g., “refugees flee” and “refugees ran away”. To ensure the reliability and comparability of created schemas, our system grounds the nodes to an ontology, namely the XPO ontology, in the last stage. Each node in the XPO ontology contains a unique node ID, a node name, and a concise description (definition), and a list of similar nodes. Our system offers two ways of grounding, ``name inference grounding" or ``name similarity grounding". Name inference grounding maps the schema nodes to XPO nodes by predicting the XPO node’s name;
% implemented by \add {citation to Zoey's grounding paper} ; 
name similarity grounding finds the XPO nodes by comparing the similarities between the embeddings of a schema node and a XPO node’s name.

In name inference grounding, given a graph node, our system first use few-shot GPT-3 to deduce a list of possible XPO names (see few-shot prompt example in appendix \ref{sec:appendix}). Then, the candidate XPO names are postprocessed by dropping off the wrong prediction and adding similar XPO names to the true prediction. After that, each possible XPO name will be checked for entailment with the original graph node. The entailment model is a BART-large model fine-tuned on the MNLI dataset \cite{lewis2020bart, williams2018broad}. The input is the original graph node as the premise and the possible XPO name as the hypothesis, and the output is the entailment score. We sort the possible XPO node names by their entailment scores. Users can view and choose from the top-$k$ suggested XPO nodes for the grounding of the original graph node. In name similarity grounding, the top-$k$ related XPO nodes are retrieved by computing the cosine-similarity of the GloVe embedding between the graph node and the name of XPO nodes \cite{pennington2014glove}. The above two methods are complementary to each other especially when users cannot find expected XPO nodes with one method. Human-curated data is saved in the backend database. A screenshot of node grounding can be seen in Figure \ref{case_study_ground}.

\section{Evaluation}
\label{sec: Evaluation}
\vspace{-1mm}
\subsection{A Case Study}
\vspace{-1mm}
% (model 在那些地方容易犯错，human和automatic)
In this section, we walk through the whole process of creating a toy schema with our interface which is much simpler than a fully developed schema. We assume the scenario is `cyber attack'.
 
\begin{figure}[h]
\vspace{-2.5mm}
\includegraphics[width=.4\textwidth]{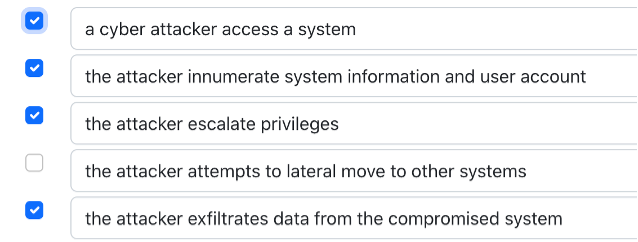}
\centering
\caption{A sample of generated steps after human-curation for scenario `cyber attack'.}
\centering
\label{case_study_step}
\vspace{-2mm}
\end{figure}

In the step generation stage, users can form a prompt from templates such as "list the steps involved in a cyber attack" with 'cyber attack' as the name and sub-event as the prompt type. Then, GPT-3 will generate 5 steps. For example, "1. A cyber attacker gains initial access to a system" and "5. The attacker exfiltrates data from the compromised system." Users can modify the content and choose steps to save. For example, one may change the first step to "1. A cyber attacker access a system." and save the step. See a screenshot of five steps for reference in figure \ref{case_study_step}.

Next, in the node extraction stage, GPT-3 will be prompted to extract nodes from the selected steps. For example, GPT-3 will output \{cyber attacker, access, system\} for the first step. The user can change the outputs to correct any mistakes. In this sample, we extract 4 nodes, they are: \{cyber attacker, access, system\}, \{attacker, enumerate, system information and user account\}, \{attacker, escalates, privileges\}, \{attacker, exfiltrate, data\}. And we concatenate the \{subject, verb, object\} into a piece of text as a node for the next stage.
 
\begin{figure}[h]
\vspace{-3mm}
\includegraphics[width=.45\textwidth]{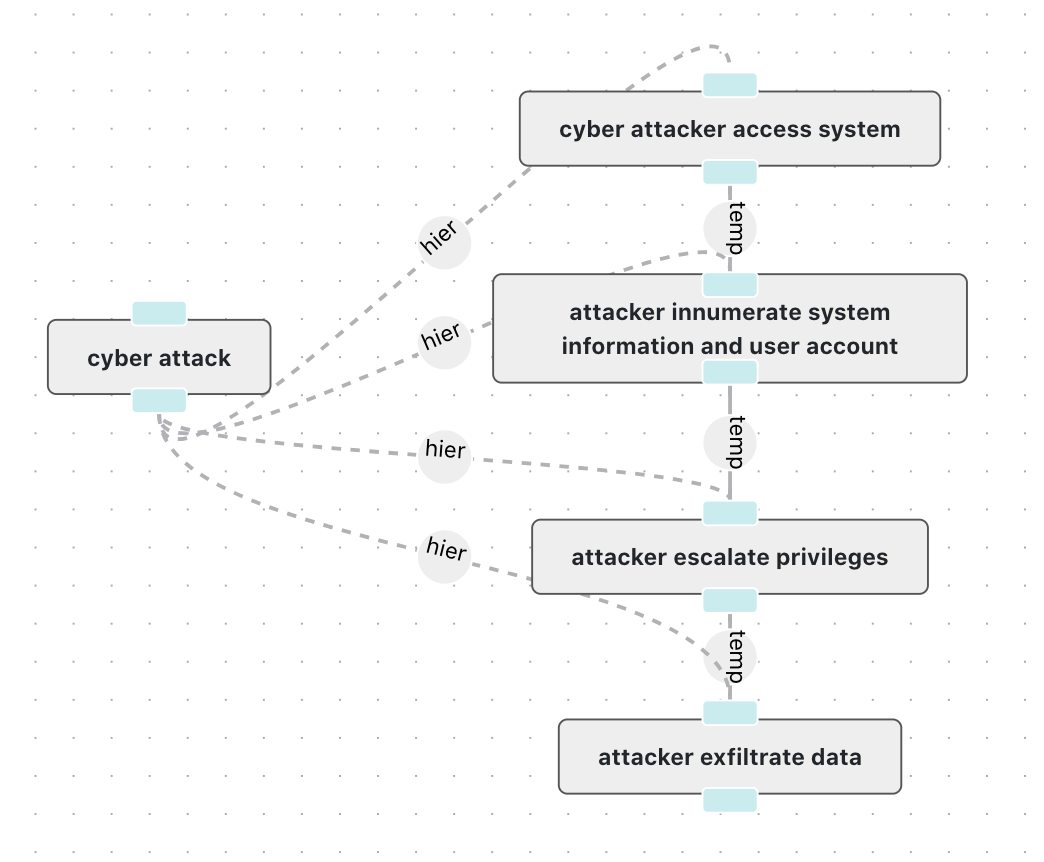}
\centering
\caption{A sample of a constructed graph after human-curation for scenario 'cyber attack'}
\centering
\label{case_study_graph}
\vspace{-1mm}
\end{figure}

Thereafter, in the graph construction stage, we prompt GPT-3 to automatically build linear temporal edges on the above four nodes that users can modify. We manually add a scenario node ‘cyber attack’ and link with the other four existing nodes through hierarchical edges. see a screenshot of the graph in figure \ref{case_study_graph}.

\begin{figure}[h]
\vspace{-1mm}
\includegraphics[width=.45\textwidth]{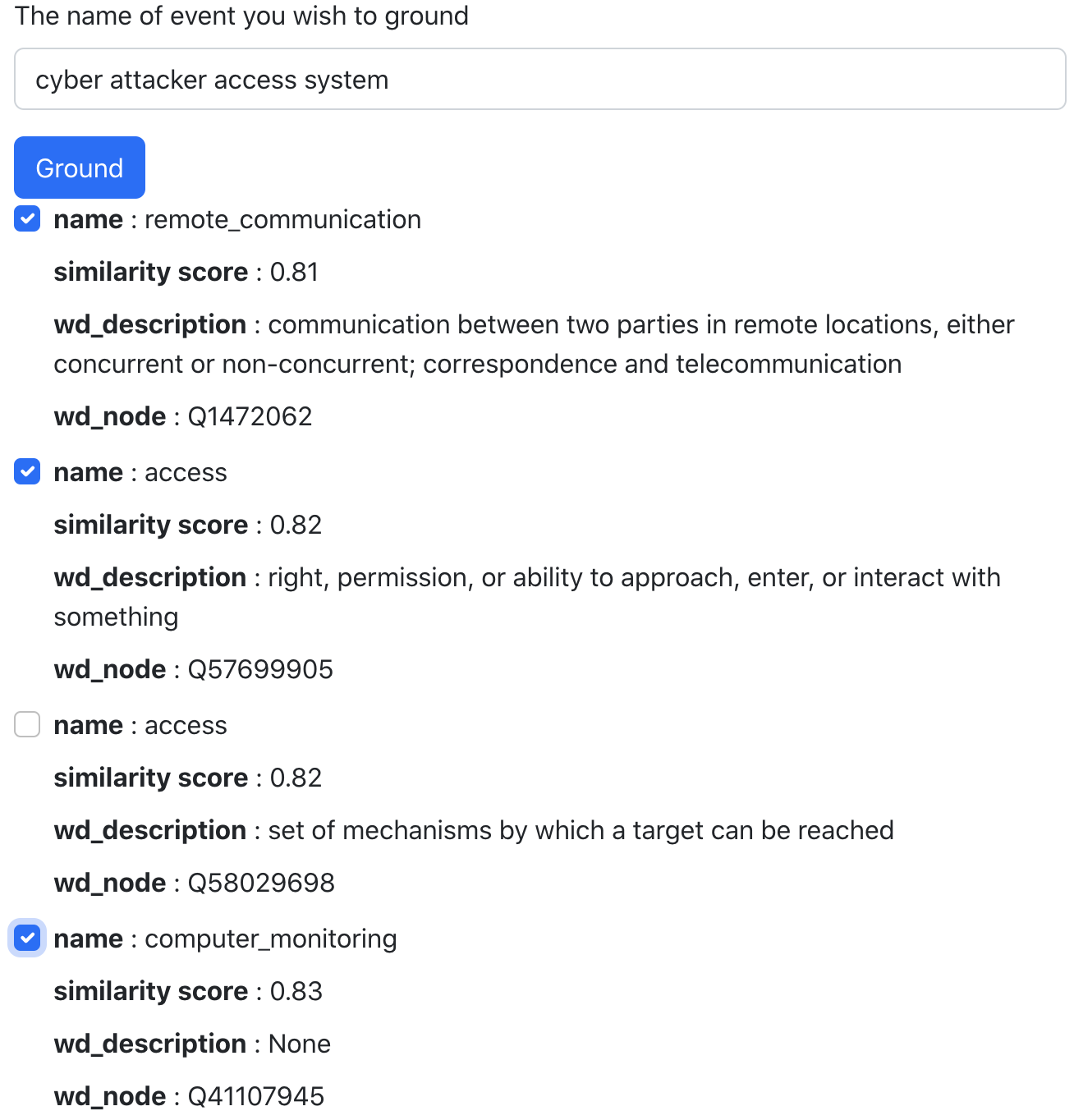}
\centering
\caption{Top-4 XPO node choices of graph node "cyber attacker access system".}
\centering
\label{case_study_ground}
\vspace{-4mm}
\end{figure}

Finally, we can optionally ground our graph node into the XPO ontology. For example, the node “cyber attacker access system” can be mapped to choices of ‘access’, ‘computer monitoring’, ‘remote communicating’ using name similarity grounding. In this case, we don't get any results from name inference grounding. 
See a screenshot of grounding in Figure \ref{case_study_ground}.

% \vspace{-1mm}
\subsection{User Evaluation}
% \vspace{-1mm}
\begin{table}
\vspace{-2.5mm}
\small
\centering
\resizebox{.48\textwidth}{!}{
\begin{tabular}{c|c|c|c|c|c}
\hline
% \textbf{Output} & \textbf{natbib command} & \textbf{Old ACL-style command}\\\hline
& EVC & FOD & JOB & MED & MRG\\\hline
Step Acc & 11/12 & 7/8 & 10/10 & 10/10& 12/12 \\\hline
Node Acc & 13/15 & 10/10 & 11/12 & 12/12 & 12/14 \\\hline
\makecell[c]{Graph Node\\ED} & 1 & 0 & 0 & 0 & 0 \\\hline
\makecell[c]{Graph Edge\\ED} & 8 & 0 & 7 & 3 & 16\\\hline
\makecell[c]{Grouding\\Success Rate}& 5/12 & 3/10 & 3/11 & 6/12 & 9/12 \\\hline
\makecell[c]{Self-reported\\time (min)} & 15 & 10& 11 & 10& 14\\\hline
\end{tabular}
}
\caption{\label{user-eval}
User evaluation results. Acc in line 1 and 2 represents Accuracy. ED in line 3 and 4 means Editing Distance.
}
\vspace{-3mm}
\end{table}
We followed the evaluation methodology used by \citet{ciosici-etal-2021-machine} with slight modifications to assess our system. Evaluation is done by researchers in the field of NLP who have experience in hand-writing event schemas but have not used the interface before. In the step generation and node extraction stage, we count the number of human selected steps/nodes out of the total number of machine generated results as accuracy. For simplicity, we ignore users’ modifications (e.g. rephrasing) at this point. In the graph construction stage, we compare how many nodes and edges are modified (added or deleted) using graph edit distance. In the grounding part, the success rate is measured as successful retrieval of at least one relevant XPO node within top-3 grounding results for a given event node. We also ask users to self-report their total time of interaction. For all the evaluations, we use GPT-3 Davinci model as the language model.\footnote{\url{https://platform.openai.com/docs/models/gpt-3}}

We follow prior work and evaluate our system on five scenarios: Evacuation (EVC), Ordering Food in a Restaurant(FOD), Finding and Starting a New Job (JOB), Obtaining Medical Treatment (MED), Corporate Merger or Acquisition (MRG).\footnote{\label{eval} Detailed evaluation results: \url{https://joeyhou.notion.site/Human-in-the-Loop-Schema-Induction-Interface-Logs-1eb52403b05542919ccea214656f4211}}

As shown in Table \ref{user-eval}, our interactive system shows high accuracy in step and node generation phases, thanks to the richness of world knowledge from LLMs. However, the graph construction and the node grouding require more human curation, due to the difficulty of event reasoning, such as the understanding of temporal and hierarchical relationships; and the  retrieval ability from large database. In those cases, we showed that human curation can step in timely and improve the quality of event schema when LLM-based models make mistakes\ref{eval}. In addition, our interface is easy to use, with much shorter times required to complete each event schema task compared to previous work \cite{ciosici-etal-2021-machine}.

% In addition to the quantitative results, 
We also observed that the node extraction results are ambiguous when the original sentence include complex attributes such as location, condition, or other modifiers. For example, given the step \textit{"waitress bring order to the kitchen"}, current node extraction produces node \textit{"(waitress, bring, order)"}, while the location information \textit{"kitchen"} is lost. A more informative node representation would be an extension in our future work. 

\section{Conclusion}
\label{sec: Conclusion}
We propose a human-in-the-loop schema curation interface with pre-trained large language models (LLMs) as the backbone. We use LLMs to generate candidate components of a schema
% , extract events from generated text, 
and involve human as the final judge for both the content and structure of the event schema. With empirical evaluations, we show that our system can efficiently produce human-validated event schemas with minori human efforts.% and accuracy.

\section*{Limitations}
We have several limitations in our current approach. First, our current system uses zero-shot or few-shot to prompt GPT-3 without any fine tuning. In future work, we plan to fine-tune our GPT-3 with human curated data that we collect.  We expect that fine-tuning will improve our models' performance. It may also be possible to use human curated data to train a policy network recommended by OpenAI \cite{ouyang2022training}. Second, we can replace GPT-3 with more robust task specific models at some stages, e.g., the pre-trained model for predicting temporal and hierarchical orders. Third, some users suggested incorporating a graph view at the other three stages, which will help users to generate based on the current graph. We will include this graph view in our next version.

\section*{Ethics Statement}
To our knowledge, our back-end GPT-3 model was trained mainly on English web data, it may prefer events happen in an English environment. Furthermore, our test showed that it generated events specifically fit in American setting, for example, Miranda Rights for arrest, Democrats and Republicans in United States for election. These facts suggest GPT-3 may ignore the knowledge of non-American cultures or minority groups. In addition, currently, we only create schemas for scenarios that are reported in mainstream news media, e.g. conflict, communication. It excludes the schemas from other domains, such as biology, medicine.

\section{Acknowledgements}
This research is based upon work supported in part by the DARPA KAIROS Program (contract FA8750-19-2-1004), the DARPA LwLL Program (contract FA8750-19-2-0201), the IARPA BETTER Program (contract 2019-19051600004), and the NSF (Award 1928631). Approved for Public Release, Distribution Unlimited. The views and conclusions contained herein are those of the authors and should not be interpreted as necessarily representing the official policies, either expressed or implied, of DARPA, IARPA, NSF, or the U.S. Government. 

We thank researchers in PennNLP groups, and from other universities who gave us suggestions on the paper.

\bibliography{anthology,custom}
\bibliographystyle{acl_natbib}

\appendix

\section{Appendix: Prompt and Completion}
\label{sec:appendix}
\subsection{Appendix A: Prompt and example completions for Step Generation}
\textbf{Prompt:} List the events before an attack: 1.\\
\textbf{Completion:}\\
The attacker gathers information about the target.\\
2. The attacker plans the attack.\\
3. The attacker gains access to the target system.\\
4. The attacker executes the attack.\\
5. The attacker covers their tracks.\\

\noindent\textbf{Prompt:} List the events after an attack: 1.\\
\textbf{Completion:}\\
The attacker's identity is confirmed.\\
2. The target is notified of the attack.\\
3. The attacker is placed on a watch list.\\
4. The attacker's device is seized.\\
5. The attacker is arrested.\\

\noindent\textbf{Prompt:} List the sub-events involved in an attack: 1.\\
\textbf{Completion:}\\
Identify the target.\\
2. Plan the attack.\\
3. Choose the weapons.\\
4. Assemble the team.\\
5. Launch the attack.\\
6. Evaluate the results.\\

\subsection{Appendix B: Prompt and example completions for Node Extraction}
\textbf{Prompt:}\\
\slash{*} instruction *\slash{}\\
For each sentence, extract event verbs and their arguments, categorizing the arguments as subject or object. Write None if there is no object.\\
Return in [verb: \_, subject: \_, object: \_] format.\\

\noindent\slash{*} few-shot examples *\slash{}\\
For example:\\
Q: Isaac ate a cake today and he played football.\\
A: [verb: eat, subject: Isaac, object: cake], [verb: play, subject: Isaac, object: football]\\

\noindent Q: The teacher arrived in class and he started teaching.\\
A: [verb: arrive, subject: teacher, object: class], [verb: start, subject: teacher, object: teaching]\\

\noindent Q: Nate and Isaac ate dinner.\\
A: [verb: eat, subject: Nate and Isaac, object: dinner]\\

\noindent Q: Justin slept.\\
A: [verb: sleep, subject: Justin, object: None]\\

\noindent\slash{*} target example *\slash{}\\
Q: The attacker gathers information about the target.\\
A:\\
\textbf{Completion:}\\
{[verb: gather, subject: attacker, object: information]}\\

\noindent\slash{*} target example *\slash{}\\
Q: The attacker's identity is confirmed.\\
A:\\
\textbf{Completion:}\\
{[verb: confirm, subject: attacker's identity, object: None]}\\

\noindent\slash{*} target example *\slash{}\\
Q: The attacker is placed on a watch list.\\
A:\\
\textbf{Completion:}\\
{[verb: place, subject: attacker, object: watch list]}

\subsection{Appendix C: Prompt and example completions for Node Grounding}
\textbf{Prompt:}\\
\noindent\slash{*} few-shot examples *\slash{}\\
List event names related to the event "People are infected with this disease":\\
1.infection\\
2.epidemic\\
3.pandemic\\

\noindent List event names related to the event "It was a robbery-related incident":\\
1.robbery\\
2.burglary\\
3.theft\\

\noindent List event names related to the event "The first case of the disease have detected and it has been reported":\\
1.infection\\
2.epidemic\\
3.pandemic\\

\noindent List event names related to the event "The disease is eventually brought under control":\\
1.control\\
2.improvement\\

\noindent List event names related to the event "People who are ill have serious symptoms":\\
1.symptoms\\

\noindent List event names related to the event "The pathogen begins to spread through the population":\\
1.transmission\\
2.spread\\

\noindent\slash{*} target example *\slash{}\\
\noindent List event names related to the event "The attacker gathers information about the target":\\
\textbf{Completion:}\\
1.reconnaissance\\
2.surveillance\\
3.investigation\\

\noindent\slash{*} target example *\slash{}\\
\noindent List event names related to the event "The attacker's identity is confirmed":\\
\textbf{Completion:}\\
1.identification\\
2.confirmation\\

\noindent\slash{*} target example *\slash{}\\
\noindent List event names related to the event "The attacker is placed on a watch list":\\
\textbf{Completion:}\\
1.surveillance\\
2.monitoring\\
3.investigation\\

\end{document}